\pdfoutput=1
\documentclass[manuscript]{aastex}

\slugcomment{Submitted to ApJ 04-22-16; revised 09-16-16; accepted 09-17-16}

\shorttitle{WISEA 0615-1247}
\shortauthors{Fajardo-Acosta et al.}

\begin{document}


\title{Discovery of a Possible Cool White Dwarf Companion from the AllWISE Motion Survey}


\author{Sergio B. Fajardo-Acosta, J. Davy Kirkpatrick}
\affil{Infrared Processing and Analysis Center, California Institute of 
Technology, \\
    Mail Code 100-22, Pasadena, CA 91125}
\email{fajardo@ipac.caltech.edu, davy@ipac.caltech.edu}

\author{Adam C. Schneider, Michael C. Cushing}
\affil{University of Toledo, 2801 W. Bancroft St., MS 113, \\
    Toledo, OH 43606}
\email{aschneid10@gmail.com, michael.cushing@utoledo.edu}

\author{Daniel Stern}
\affil{Jet Propulsion Laboratory, California Institute of Technology, \\
    4800 Oak Grove Dr., MS 169-221, Pasadena, CA 91109}
\email{daniel.k.stern@jpl.nasa.gov}

\author{Christopher R. Gelino}
\affil{Infrared Processing and Analysis Center \& NASA Exoplanet Science Institute, \\ California Institute of 
Technology, \\
    Mail Code 100-22, Pasadena, CA 91125}
\email{cgelino@ipac.caltech.edu}

\author{Daniella C. Bardalez-Gagliuffi}
\affil{University of California at San Diego, \\ 
    9450 Gillman Dr. \# 40282, La Jolla, CA 92092}
\email{daniella@physics.ucsd.edu}

\author{Kendra Kellogg}
\affil{Western University, 226-376-3530, 454 Castlegrove Blvd., \\
    London, ON N6G 1K8, Canada}
\email{kkellog@uwo.ca}

\and

\author{Edward L. Wright}
\affil{University of California at Los Angeles, \\
    Dept.\ of Physics \& Astronomy, P.O. Box 951547, \\
    Los Angeles, CA 90095-1547}
\email{wright@astro.ucla.edu}

\begin{abstract}
We present optical and near-infrared spectroscopy of 
WISEA J061543.91$-$124726.8, which we rediscovered as a high motion 
object in the AllWISE survey. 
The spectra of this object 
are unusual; while the red optical ($\lambda >$ 7,000 \AA) and near-infrared
spectra exhibit characteristic TiO, VO, and H$_{2}$O bands
of a late-M dwarf, the blue portion of its optical spectrum shows a 
significant excess of emission relative to late-M type templates. The
excess emission is relatively featureless, with the exception of a 
prominent and very broad Na I D doublet. We
find that no single, ordinary star can reproduce these spectral 
characteristics. The most likely explanation is an unresolved 
binary system of an M7 dwarf and a cool white dwarf. The flux of a cool 
white dwarf drops in the optical red and near-infrared, 
due to collision-induced absorption,
thus allowing the flux of a late-M dwarf to show through. This scenario,
however, does not explain the Na D feature, which is unlike that of 
any known white dwarf, but which could perhaps be explained via 
unusual abundance or pressure conditions.

\end{abstract}

\keywords{brown dwarfs, solar neighborhood, stars: fundamental parameters, stars: low-mass, subdwarfs, white dwarfs, stars: individual (WISEA 0615$-$1247)}

\section{Introduction}\label{intro}

The study and census of our Solar neighborhood is important to advance
our understanding of low mass stars and brown dwarfs, which are the most 
numerous known objects in the Galaxy. As such they shed light on the 
low-end
of the initial mass function, and on star formation efficiency. The search
for our nearest neighbors lets us study objects in great detail, 
because they are the closest, brightest objects of their class. 
The detailed study of late-type dwarfs, brown dwarfs,
subdwarfs, and white dwarfs has historically relied on the identification 
of objects based on their optical through mid-infrared colors. These studies
have been enhanced by recent surveys for high proper motion objects, 
because 
these surveys have no color bias, and can therefore identify objects with 
unusual characteristics not found through color-based searches.

\begin{figure}
\includegraphics[angle=90, width=6.5in]{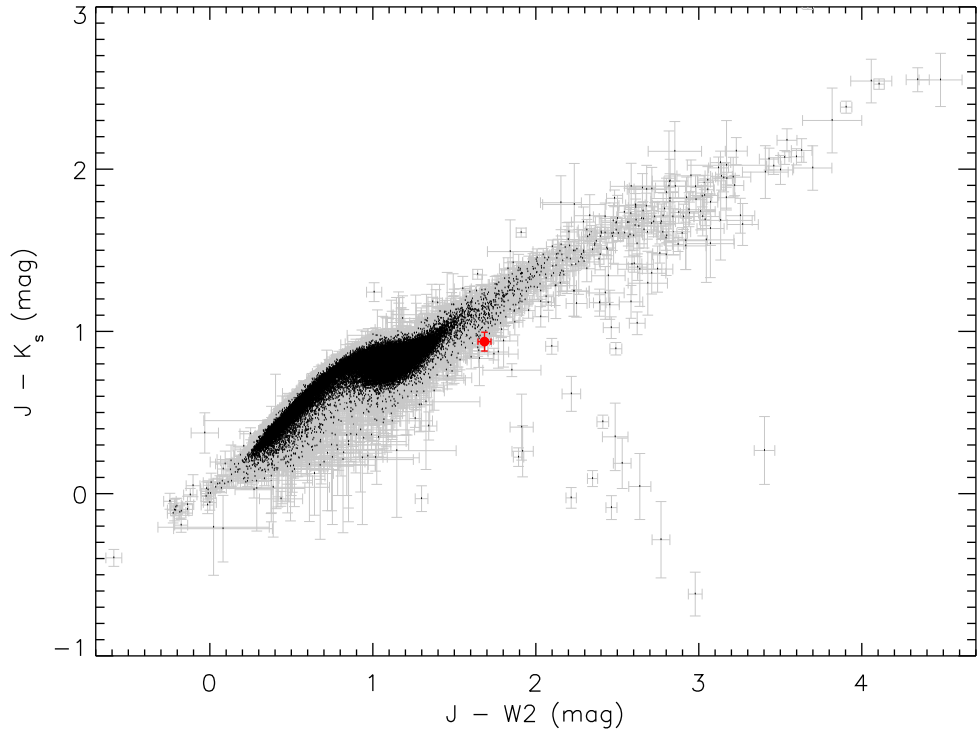}
\caption{Color $J - K_{s}$ plotted as a function of color $J - W2$ for
47,936 high motion objects ({\it black symbols with gray error bars}) 
from Kirkpatrick et al.\ (2014) and Kirkpatrick et al.\ (2016).
The {\it red filled circle} marks the location of 
WISEA 0615-1247. \label{fig1}}
\end{figure}

The object 2MASS J06154357$-$1247221 = WISEA J061543.91$-$124726.8, which we 
henceforth designate WISEA 0615$-$1247, was originally discovered as a high
proper motion star by L\'{e}pine (2008) and designated by him as 
PM I06157$-$1247 in a re-analysis of southern hemisphere 
digitized sky surveys. L\'{e}pine (2008) finds proper motion components for
WISEA 0615$-$1247 of $\mu_{\alpha} =$ 452 $\pm$ 10 mas\, yr$^{-1}$, and 
$\mu_{\delta} = -$421 $\pm$ 10 mas\, yr$^{-1}$ from measurements spanning 
an epoch range of 39.2 years (see Table 1). The object was rediscovered 
by Kirkpatrick et al.\ (2014) due to its high motion in the AllWISE Data 
Release of the Wide-Field
Infrared Survey Explorer ({\it WISE}; Wright et al.\ 2010).
The colors of the object, namely $J - K_{s} =$ 0.937 $\pm$ 0.058 mag, from 
the Two Micron All-Sky Survey (2MASS; Skrutskie 1997), and 
$J - W2 =$ 1.685 $\pm$ 0.040 mag (from 2MASS $J$ and the {\it WISE} 
4.6 $\mu$m bandpass $W2$; see Table 1), are shown as the red symbol in 
Figure~\ref{fig1}.
This figure compares these colors with those of 47,936 high motion sources 
(Kirkpatrick et al.\ 2014, 2016).  
WISEA 0615$-$1247 lies $\sim$ 0.2 mag below or blueward in $J - K_{s}$ 
color relative to the
normal sequence of field objects, possibly hinting at a cool subdwarf
nature. Early-L subdwarfs have colors $J - W2$ ranging from 1.1 to 1.8 
mag, and $J - K_{s}$ on average 0.23 mag below the normal sequence of field
objects (see Figure 7 in Kirkpatrick et al.\ 2016). 
The suggested sub-dwarf classification of the object is similar to that by 
L\'{e}pine (2008),
based on the reduced proper motion of the object in the $V$ bandpass. 
In order to elucidate
the nature of the object, we obtained optical and near-infrared spectroscopy. 
The spectrum, as discussed below, proved to be unique. We propose a binary
scenario that we believe best explains the observed data, although it
invokes a cool white dwarf companion unlike any heretofore cataloged.

\section{Observations}


\subsection{Proper Motion} \label{obs_proper_motion}

The AllWISE Source Catalog\footnote{See http://wise2.ipac.caltech.edu/docs/release/allwise/expsup/. Motions listed in the AllWISE source Catalog are not
strictly proper motions, because parallax effects are not taken into account.
Hence we list them as ``RA Motion'' and ``Dec Motion,'' to differentiate
them from proper motions from other catalogs, which
we list as ``$\mu_{\alpha}$'' and ``$\mu_{\delta}$.''} lists 
RA Motion = 593 $\pm$ 73 mas\, yr$^{-1}$ and 
Dec Motion = $-$322 $\pm$ 78 mas\, yr$^{-1}$ for WISEA 0615$-$1247 (Table 1).
In order to confirm this motion, we measured the displacement of the object
between the 2MASS and AllWISE positions, 
to yield $\mu_{\alpha}$ = 436.9 $\pm$ 6.4 mas\, yr$^{-1}$ and 
$\mu_{\delta}$ = $-$406.8 $\pm$ 6.5
mas\, yr$^{-1}$ (Table 1). This proper motion is in rough agreement, 
within the errors, 
with the AllWISE motions. We carried out an additional estimate of the 
motion of 
the object, 
via a $\chi^{2}$ minimization fit to its positions in the First
and Second Digitized Sky Surveys (DSS1 and 
DSS2)\footnote{See http://archive.stsci.edu/dss/} red ($R$-band) plates, 
2MASS, {\it WISE} All-Sky\footnote{See http://wise2.ipac.caltech.edu/docs/release/allsky/expsup/.}, and {\it WISE} 3-Band Cryo\footnote{See http://wise2.ipac.caltech.edu/docs/release/allsky/expsup/section7\_1.html.} Data Releases. In this way we 
obtained $\mu_{\alpha}$ = 447 $\pm$ 5 and $\mu_{\delta}$ = $-$414 $\pm$ 4 mas\, yr$^{-1}$, as 
well as a parallax $\pi$ which is practically 0 due to the large errors 
listed in table 1. The proper
motion of the object thus dominates the AllWISE measured motion.

\subsection{Optical Spectroscopy}\label{optical_spectra}

We observed the object with the Double Spectrograph (DBSP; Oke \& Gunn 1982) 
on the Hale 
200-inch Telescope at Palomar Observatory, on UT 2014 Feb 23, Sep 27, 
Oct 24, and Nov 15. On the blue side of DBSP, a grating with 600 lines/mm
blazed at 4,000 \AA\ was used for all nights. On the red side a grating with 
600 lines/mm
blazed at 10,000 \AA\ was used, except on Nov 15, when a red side
grating with 316 lines/mm blazed at 7,500 \AA\ was used. The D68 dichroic,
which splits the light between the blue and red sides near 6,800 \AA, was
used on most nights, except on Nov 15, when the D55 dichroic was used, which
splits the light near 5,500 \AA. The resulting spectra 
covered the range 4,100 to 9,900 \AA, except that the spectrum on Nov 15 
covered from 2,900 to 10,640 \AA. Grating angles were selected for
each dichroic in order to provide wavelength overlap between the blue and 
red sides of the spectra. Typical spectral resolution with the 1.0 arcsec
slit was $\sim$ 3.6 \AA\ in the blue, and $\sim$ 2.4 \AA\ in the red. 
The integration times with DBSP
ranged from 3,360 sec on Oct 24 to 6,060 sec on Nov 15. Conditions were clear
on Feb 23; windy with thin clouds on Sep 27; clear with 0.8\arcsec\ seeing on
Oct 24; and some cloudiness on Nov 15.

We also observed the object with the Low Resolution Imaging 
Spectrometer (LRIS; Oke et al.\ 1995) on the Keck I 10-meter telescope at 
the W. M. Keck Observatory, on UT 2014 Oct 20. On the blue side of LRIS, 
a grating
with 600 lines/mm blazed at 4,000 \AA\ was used, and on the red side a grating
with 400 lines/mm blazed at 8,500 \AA\ was used. The 560 dichroic was used
to split light between the blue and red sides, near 5,600 \AA. The resulting 
spectrum covered the range 
from 3,300 to 10,000 \AA. The spectral resolution with the 1.0 arcsec slit was
$\sim$ 4.6 \AA\ in the blue, and $\sim$ 8.2 \AA\ in the red. 
The integration time with LRIS was 600 sec, and 
there were some clouds as remnants from hurricane Ana. Reductions for DBSP
observations made use
of standard procedures for CCD data, described in Kirkpatrick et al.\ (2016). 
Reductions for the LRIS observation made use of a non-stardard procedure,
because weather conditions prevented the observation of a flux calibrator on
that night. Instead, a flux calibrator (the white dwarf HZ 44) was 
measured on UT 2015 Jul 17, using a different instrument configuration than
for the target. Therefore we had to force the spectral shape of the LRIS 
data of the target at 8,590--10,270 \AA\ to match that of the DBSP 2014 
Oct 24 data, by fitting second order polynomials. Figure~\ref{fig2} 
shows all measured optical spectra of WISEA 0615$-$1247.

We obtained optical spectroscopy of the M7 dwarf G59-32B, also designated 
WISEA 124007.18$+$204828.9, noted as a component in a 
common-proper motion binary by Kirkpatrick et al.\ (2014). The system was
in a pre-flare state during the observations, and was intended as a
spectral-type comparison. We used the Boller \& Chivens Spectrograph on the
Irenee du Pont 2.5-meter telescope at Las Campanas Observatory, on UT 2014
May 4. The integration time was 1800 sec. See Kirkpatrick et al.\ (2016) for
more details.

We also obtained DBSP observations of three white dwarfs for comparison with
WISEA 0615$-$1247, and another object that was initially believed to also
be a white dwarf. The white dwarfs LHS 3250 and 
LSPM J1702$+$7158N were observed on UT 2015
Jun 8, with 3,660 sec integrations each, under clear conditions but with 
2\arcsec\ seeing. The white dwarf WD J0205$-$053 was
observed on UT 2015 Sep 7, with 2,400 sec integrations, through some 
clouds and glow from
a last-quarter Moon, although with 1.1--1.2\arcsec\ seeing. The 600 lines/mm 
gratings on the
blue and red sides of DBSP were again used with the D68 dichroic, to
produce continuous 4,100--9,900 \AA\ spectra. Details on these three white 
dwarfs are given in Section~\ref{cool_white_dwarfs}. We observed the object 
LP 343-35 = WISEA 221515.51$+$315731.9 (Kirkpatrick et al.\ 2016), under the 
assumption that is was the white dwarf WD 2213$+$317. This DBSP observation 
was on UT 2016 Jan 13, with a 300 sec integration, using the D55 dichroic, 
and a 1.5-arcsec slit.
The gratings used were the 600 lines/mm on the blue side, and a 316 lines/mm
one blazed at 7,500 \AA\ on the red side, to produce a continuous 
3,700--10,300 \AA\ spectrum, with typical spectral resolution of 4.9 \AA\ in
the blue, and 3.9 \AA\ in the red. 
Details on the object 
are given in Section~\ref{cool_white_dwarfs} and in the Appendix.

\subsection{Near-infrared Spectroscopy}

We observed WISEA 0615$-$1247 in the near-infrared with the upgraded SpeX 
spectrograph (Rayner et al.\ 2003) at the NASA Infrared 3-meter Telescope 
Facility, on UT 2015 Jan 18. A low-resolution ($\lambda /\Delta \lambda =$
75--120) spectrum covering the range 0.69 to 2.54 $\mu$m was obtained using 
the prism mode. Exposures were done
in an ABBA nod pattern along the 15-arcsec slit; the total integration time
was 956 sec. In order to apply telluric corrections, an A0 V star was 
observed, as close in airmass as possible to our target, although a 0.3
airmass difference was the best we could achieve. Data reduction was
done with the SpeXtool package (Vacca, Cushing, \& Rayner 2003; 
Cushing, Vacca, \& Rayner 2004).
Figure 3 shows the SpeX spectrum of WISEA 0615$-$1247. Wavelength regions
of low atmospheric transmission (less than about 60\%) are indicated with 
gray bands. These regions correspond to water vapor absorptions at 0.92,
1.13, 1.35, and 1.85 $\mu$m, as well as CO$_{2}$ absorptions near 1.6, 2.0,
and 2.06 $\mu$m (Rayner, Cushing, \& Vacca 2009). The spectrum exhibits two 
peaks at $\sim$ 1.08 and 1.27 $\mu$m, which are near, but not at, 
the positions of H Paschen-series $\gamma$ and $\beta$ lines, respectively.
The feature at 1.08 $\mu$m is likely due to telluric H$_{\rm 2}$O, and in our
experience is sometimes easier to see at the low spectral resolution of 
these data. The feature at 1.27 $\mu$m is due to poor correction of a telluric
O$_{2}$ band.

\subsection{Previous Imaging} \label{prev_image}

We searched the literature for optical imaging data of WISEA 0615$-$1247.
The purpose was to see if a background contaminant source may have been 
present in the beam during our optical spectroscopy of the object. The 
images examined were the $B$-band and $R$-band images in DSS2, and 
the $I$-band image in the Deep Near Infrared Survey of the 
Southern Sky (DENIS, 3rd Data Release\footnote{The DENIS Project website, including access to imaging data, is at \\
http://cdsweb.u-strasbg.fr/denis.html.}).
Figure~\ref{fig4} shows these three images.

Overlaid in Figure~\ref{fig4} are the positions of WISEA 0615$-$1247,
shown as green circles and numbered labels, at respectively the $B$-band 
(1983), $R$-band (1993), $I$-band (1999), AllWISE (2010), and 
our optical
spectroscopy (2014) epochs. The latter were computed from the AllWISE 
position of the object, and our estimate of motion from the displacement 
between 2MASS and AllWISE (Section~\ref{obs_proper_motion}). The  
orange arrow overlaid on the $R$-band image shows the motion of 
WISEA 0615$-$1247. The source just below
the motion vector and to the right of the AllWISE position (number 4) is
a neighboring DENIS Catalog non-moving source. This DENIS 
source was 4 or 5 arcsec from WISEA 0615$-$1247 during our spectroscopic 
observations, and hence did not contaminate our spectrum since our slit
width was 1.0 arcsec. No other sources with strong $B$- or $R$-band flux are seen
in the field of our object at earlier epochs, hence we conclude that there
is no contamination of our spectrum by a background source.

\subsection{Search For Other Archival Observations}

There were no
X-ray measurements of this object in the {\it ROSAT} All Sky 
Survey\footnote{See the 
{\it ROSAT} All Sky Survey page at \\
http://www.xray.mpe.mpg.de/cgi-bin/rosat/rosat-survey}, or UV measurements
in {\it GALEX}\footnote{Information about {\it GALEX} is in 
http://www.galex.caltech.edu/ and http://galexgi.gsfc.nasa.gov/}, or radio 
measurements in the VLA Faint Images of the Radio Sky at Twenty-cm (FIRST)
Survey\footnote{The FIRST Survey is described in http://sundog.stsci.edu/}.

\section{Possible Explanations}

Our repeated optical spectra of WISEA 0615$-$1247 measured throughout 2014
(Figure~\ref{fig2}) do not show statistically significant evidence of
variability. The spectral slopes are very comparable among the
various observations (although the LRIS 2014 Oct 20 spectrum was modified,
as described in Section~\ref{optical_spectra}). 

\subsection{Single Object Hypothesis} \label{single_object}

In order to compare our spectra of WISEA 0615$-$1247 with those of M dwarfs, 
we first combined our optical and near-infrared data into a
single spectrum. We used our UT 2014 Oct 24 DBSP spectrum which has the
best signal-to-noise,
and normalized our SpeX spectrum to it at 7,916 \AA. We then trimmed the two
spectra into contiguous but non-overlapping ranges (4,033--8,764 \AA\ for 
DBSP, and 8,783--24,950 \AA\ for SpeX), taking into account that
the optical red portion of the SpeX spectrum had better signal-to-noise than
the DBSP one. The combined DBSP and SpeX spectrum is shown replicated in 
black in Figure~\ref{fig5}, and is compared with spectra of M5 through 
M9 dwarfs from the SpeX Prism Spectral Libraries\footnote{The SpeX Prism Spectral Libraries are in 
http://browndwarfs.org/spexprism/.} (Burgasser 2014), shown in red in 
Figure~\ref{fig5}.

\begin{figure}
\includegraphics[angle=0, width=5.7in]{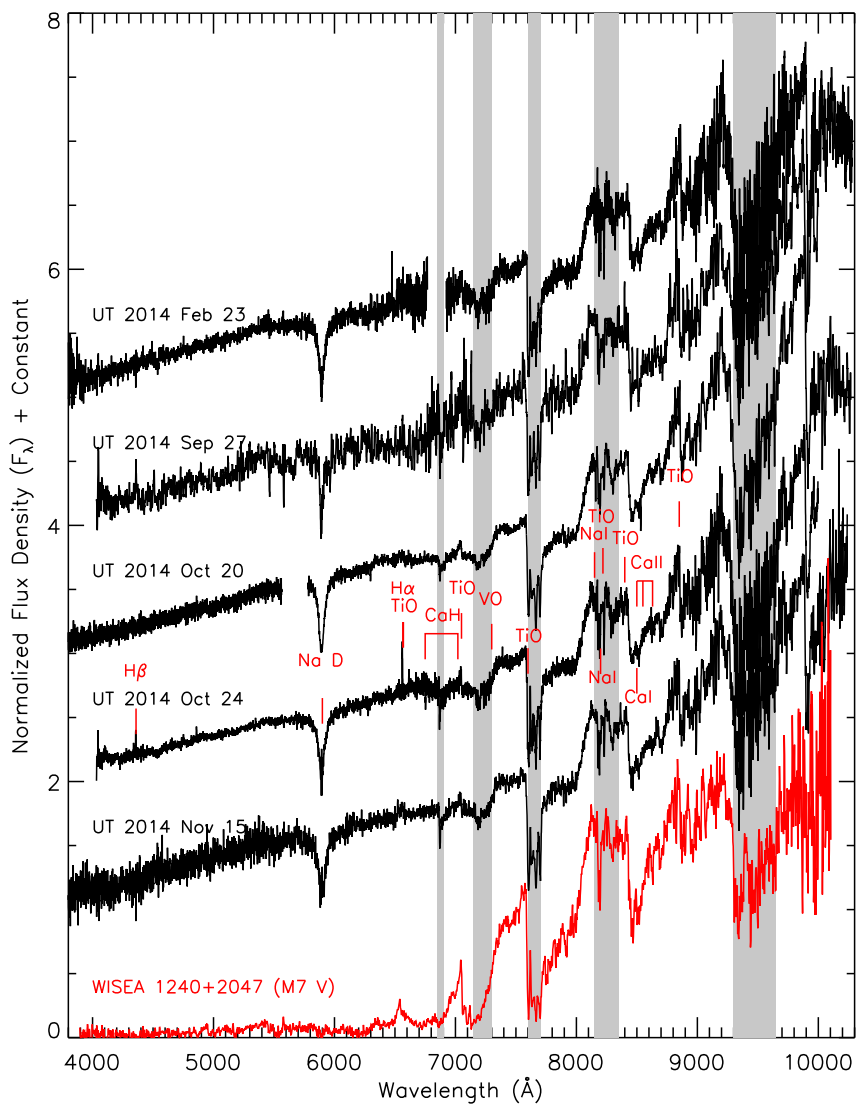}
\caption{\footnotesize Optical spectra of WISEA 0615$-$1247, measured with DBSP 
at Palomar Observatory, on UT 2014 Feb 23, Sep 27, Oct 24,
and Nov 15, and with LRIS at Keck Observatory, on UT
2014 Oct 20. The bottom spectrum, shown in {\it red}, is of the 
M7 dwarf G59-32B, or WISEA 124007.18$+$204828.9, for spectral-type 
comparison. 
Spectra have been normalized at 7,500 \AA, and a constant
offset has been added to separate the spectra vertically. Regions of low
atmospheric transmission (Hamuy et al.\ 1994) are indicated by 
{\it gray bands}. Feature identifications are shown in red for the Oct 24
spectrum of WISEA 0615$-$1247.\label{fig2}}
\end{figure}

\begin{figure}
\includegraphics[angle=90, width=6.5in]{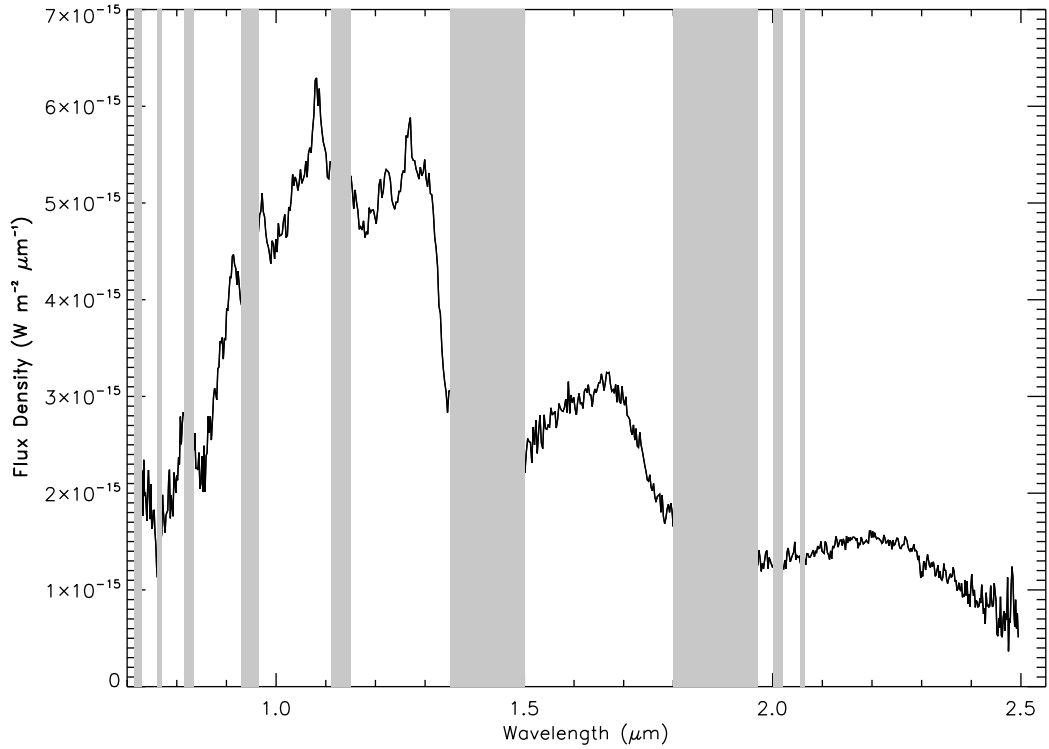}
\caption{Near-infrared spectrum of WISEA 0615$-$1247, measured with the SpeX
spectrograph at the NASA IRTF, on UT 2015 Jan 18. Regions of telluric 
absorption (Rayner, Cushing, \& Vacca 2009) are indicated by 
{\it gray bands}.\label{fig3}}
\end{figure}

\begin{figure}
\includegraphics[angle=0, width=6.5in]{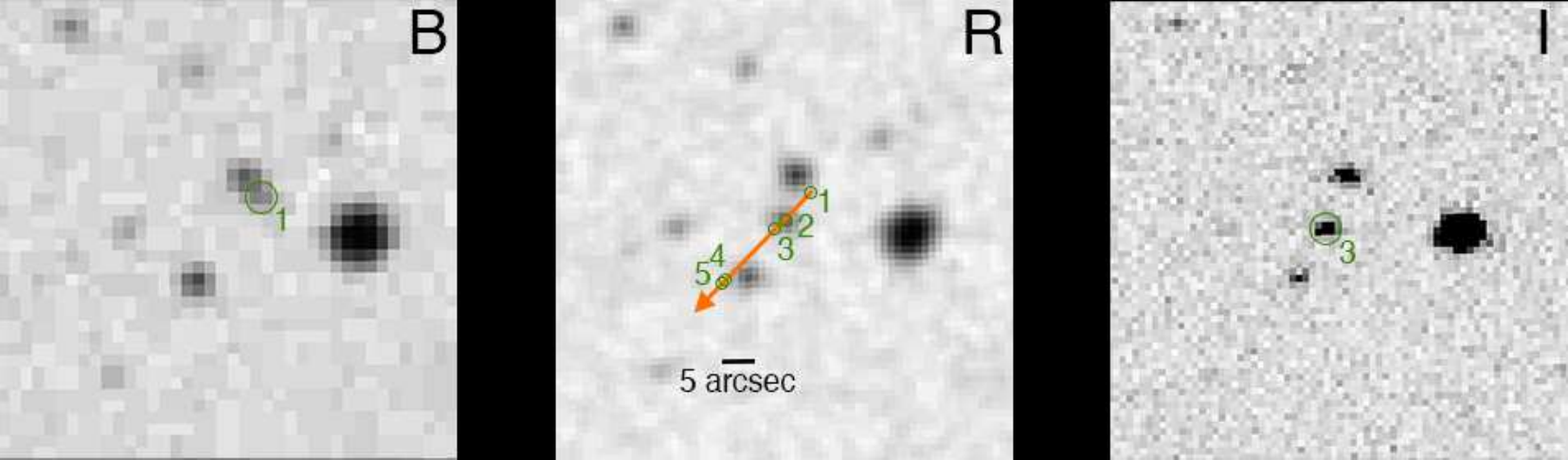}
\caption{Image cut-outs (66$\times$66 arcsec$^{2}$) showing the 
positions of WISEA 0615$-$1247 at various epochs. The {\it left panel} is the
DSS2 $B$-band image, at epoch 1983; the {\it middle panel} is the
DSS2 $R$-band image, at epoch 1993; and the {\it right panel} is 
the DENIS $I$-band image, at epoch 1999. The positions of WISEA 0615$-$1247
are indicated by the overlaid {\it green circles} and {\it numbered labels} 
1 through 5, which correspond, respectively, to the $B$-band, $R$-band, 
$I$-band, 
AllWISE, and our optical spectroscopy epochs. The {\it orange arrow} overlaid
on the $R$-band image shows the motion of WISEA 0615$-$1247 throughout these
epochs. The source just below the motion vector and to the right of the 
AllWISE position (number 4) is a neighboring DENIS Catalog 
non-moving source.\label{fig4}}
\end{figure}

\begin{figure}
\includegraphics[angle=0, width=5.5in]{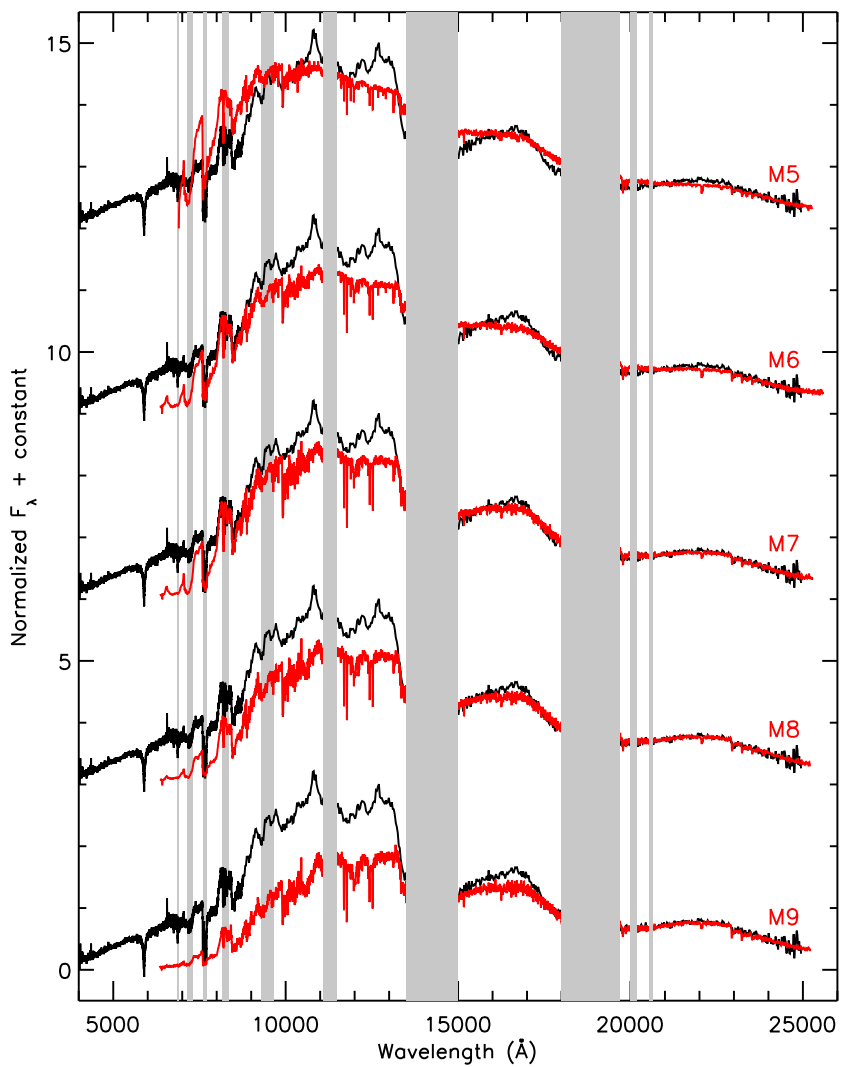}
\caption{\footnotesize The {\it black lines} are the combination of our DBSP spectrum of 
WISEA 0615$-$1247 measured on UT 2014 Oct 24, trimmed to the range 4,033 
through 8,764 \AA, and our SpeX spectrum trimmed to the range 8,783 \AA\ 
through 2.495 $\mu$m. The SpeX spectrum was normalized to the DBSP one at 
7,916 \AA, before trimming. The combined spectrum, 
in normalized flux density units, was replicated vertically by adding 
offsets, for comparison with spectra of known mid- through late-M type 
dwarfs (shown as {\it red lines}) from the SpeX Prism Spectral Libraries; see
text for reference. These comparison spectra were normalized to our object
spectrum at 2.1 $\mu$m.\label{fig5}}
\end{figure}

At wavelengths $>$ 7000 \AA, Figures~\ref{fig2} and~\ref{fig5} show that
an M7 dwarf is the best overall fit to our spectrum, among the mid- to 
late-M dwarfs considered, but with a deficit of flux at 9,000 \AA\ to 
1.3 $\mu$m. Our spectrum exhibits numerous atomic and
molecular features (Ca I, Ca II, TiO, and VO) indicating an M dwarf of 
solar metallicity 
(Figure~\ref{fig2}). The spectrum also exhibits moderate 
H$\alpha$ and H$\beta$ emission lines. The former (H$\alpha$), 
absorption 
features of VO (7,300 \AA) and TiO (8,850 \AA), and an absorption line of 
Ca I (8,500 \AA) suggest a late-M dwarf classification (Kirkpatrick 2009). 
The near-infrared spectrum (Figure~\ref{fig3}) shows pronounced absorption 
bands of H$_{2}$O, wider than telluric ones, at 1.36--1.57 and 1.72--2.1 
$\mu$m, also indicative of a late M dwarf classification (Kirkpatrick 2009).

\begin{figure}
\includegraphics[angle=0, width=5.25in]{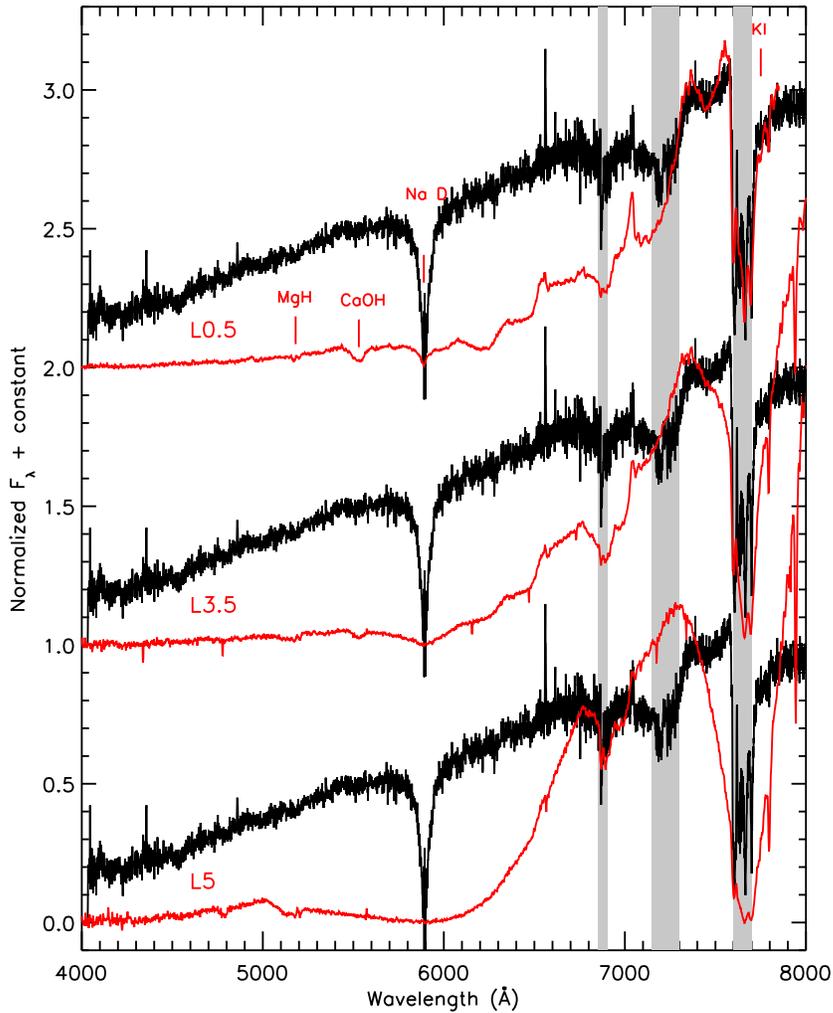}
\caption{\footnotesize The {\it black lines} are our DBSP spectrum of 
WISEA 0615$-$1247 measured on UT 2014 Oct 24. The spectrum is plotted in 
normalized flux density units, and was replicated vertically by adding 
offsets, for comparison with spectra of known early to mid-L type 
dwarfs (shown as {\it red lines}). These spectra are of 
2MASS J07464256$+$2000321 (L0.5), 2MASS J00361617$+$1821104 (L3.5), and 
2MASS J15074769$-$1627386 (L5), from Kirkpatrick et al.\ (2000) and 
Reid et al.\ (2000).
The spectra are plotted only up to 8,000 \AA\ because of large discrepancies
with our object at longer wavelengths. Regions of low atmospheric 
transmission (Hamuy et al.\ 1994) are indicated by 
{\it gray bands}. Feature identifications are shown in red for the L0.5
dwarf spectrum. \label{fig6}}
\end{figure}

In the blue optical, 
4,000--7,200 \AA, the spectral slopes of the M7 dwarf shown in 
Figure~\ref{fig2}, or M dwarfs shown in Figure~\ref{fig5}, do not 
fit the spectral slope of our object. There is a prominent Na D absorption
line at 5,900 \AA\ in our object, that is not comparable in its intensity and
width to that in M or L dwarfs.
An early L-type dwarf spectrum would exhibit Na D 
absorption, although broader and of very different morphology 
than in our 
object. It would also exhibit features not seen in the spectrum
of our object, most notably a prominent absorption feature from a K I doublet
at 7,750 \AA, or absorption features from MgH (5,200 \AA) and 
CaOH (5,500 \AA), among others (Kirkpatrick 2009); see Figure~\ref{fig6}.

The red optical spectra of WISEA 0615$-$1247 and an M7 dwarf are 
very different from 
those of late (M)-type subdwarfs (sdM), extreme subdwarfs (esdM), and ultra 
subdwarfs (usdM). The 7,800-8,100 \AA\ spectral slope of an sdM8--9
is more smoothly monotonic in comparison with an M7 dwarf 
(Kirkpatrick et al.\ 2016;
see their Figures 62 and 39). The corresponding slope of an esdM8--8.5 or
a usdM8.5--9 is much bluer than for an M7 dwarf 
(Kirkpatrick et al.\ 2016; see their Figures 71 and 74). Those same
Figures show that the 8,500--9,250 \AA\ spectral region is very red in an 
M7 dwarf, while it is modestly red or flat in an sdM8--9, and
turns over blue with increasing wavelength in an esdM8--8.5, or usdM8.5--9 
(Kirkpatrick et al.\ 2016). 
The spectra of sdM8--9, esdM8--8.5, or usdM8.5--9 exhibit a very dramatic 
pseudo-continuum bump or CaH absorption trough at 6,800 \AA, in 
contrast to a barely noticeable feature in an M7 dwarf
(Kirkpatrick et al.\ 2016), owing to TiO bands that swamp the CaH trough
in the latter.
Therefore a late-type sdM, esdM, or usdM cannot
reproduce the red optical spectrum of WISEA 0615$-$1247 while a solar 
metallicity M7 dwarf does.


\subsection{Binary Hypothesis}\label{binary}

In this section we explore the possibility that an unresolved binary 
fits the data best. The binary would be comprised of an object 
responsible for 
the blue optical portion of the spectrum, while a late-M dwarf is 
largely responsible for the red optical and near-infrared portions.

\begin{figure}
\includegraphics[angle=90, width=6.5in]{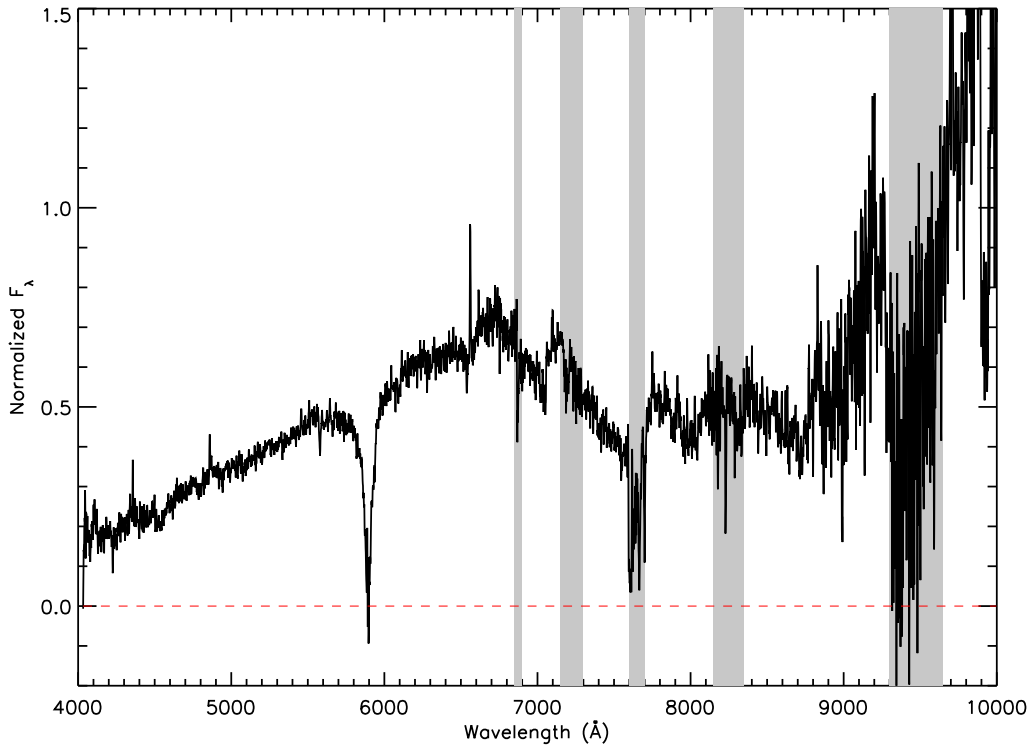}
\caption{Difference between our 2014 Oct 24 spectrum 
of WISEA 0615$-$1247, and the spectrum of the M7 V star WISEA 1240$+$2047 
(both shown in Figure~\ref{fig2}). The M7 V star was normalized to 55\% of
our object's flux density at 7,400 \AA, before subtraction, as explained
in Section~\ref{binary}. The resulting
difference has been smoothed with a 3-pixel boxcar. The {\it red dashed line}
indicates a zero difference. Regions of low atmospheric transmission 
(Hamuy et al.\ 1994) are indicated by {\it gray bands}.\label{fig7}}
\end{figure}

The spectral energy distribution of a possible binary companion to a
late-M dwarf can be assessed by subtracting a spectrum of the latter, from
that of WISEA 0615$-$1247. Our goals for the difference spectrum were for 
it to be positive in the optical, except in telluric regions, and to 
eliminate the CaH and TiO spectral features at 6,750--7,050 \AA, under
the assumption that they are produced only by the late-M dwarf. 
The one parameter that we varied was
the normalization factor of the late-M dwarf spectrum at a fixed wavelength,
which then determined the corresponding normalization factor of the
binary companion spectrum. We chose 7,400 \AA\ as the normalization
wavelength, because it is not in a telluric region, and both the spectrum
of our object and a late-M dwarf are relatively featureless thereat.
After experimenting with various normalization factors, we chose to
normalize the late-M dwarf spectrum to 55\% of the flux of our object at
the above wavelength. 
Figure~\ref{fig7} shows the subtraction of the 
M7 dwarf spectrum in Figure~\ref{fig2}, from our 
object's DBSP optical spectrum of 2014 Oct 24. 
Figure~\ref{fig7} shows that the CaH and TiO features were
not eliminated, but our choice of normalization minimized their intensity.
For normalization factors below 55\%, the above CaH and TiO features were
more prominent, and for normalization factors above 55\%, negative fluxes
resulted in various wavelength regions apart from telluric ones.
Figure~\ref{fig7} shows that
the spectrum of a possible companion to a late-M dwarf linearly rises
from 4,000 to 6,700 \AA, except that it exhibits a broad absorption 
feature of Na I D at 5,900 \AA. The spectrum then precipitously drops 
from $\sim$ 6,700 to 7,600 \AA. It is very roughly flat from $\sim$ 7,700 to
8,800 \AA, and finally rises in the remaining reddest part of it.

The difference spectrum in Figure~\ref{fig7}, and particularly the excess in 
the optical
blue, is plausibly produced by an object hotter than an M dwarf, and of 
relatively small radius, for its flux to be of the same order of magnitude as
the M dwarf. Therefore we considered a cool white dwarf as a possibility.
Cool white dwarfs are those with temperatures $\lesssim$ 4,500 K. They
are sometimes uncovered in proper motion surveys (Scholz et al.\ 2009, and 
references therein). Their 
infrared flux densities are depressed due to collision-induced absorption 
(CIA) by H$_{2}$ 
(Saumon et al.\ 1994; Hansen 1998; Abel et al.\ 2011). 
We considered CIA mainly from H$_{2}$-H$_{2}$ pairs, such as in the 
models by Abel et al.\ (2011). However, depending on the relative abundance of
He, CIA from H$_{2}$-He pairs may also be significant, as in models by
Borysow et al.\ (1997).
The effect of CIA is most evident
at wavelengths $>$ 1 $\mu$m. However, Oppenheimer 
et al.\ (2001) noted that the optical colors of cool white dwarfs are also
affected by CIA. The opacity of CIA starts at approximately 
0.6 $\mu$m and increases
towards the near-infrared (Borysow et al.\ 1997). Models of cool white dwarf 
spectra with CIA, by D.\ Saumon and D.\ Koester (presented by 
Kilic et al.\ 2006), show that their SEDs are bluer
and depressed relative to those of blackbodies. These models show that
the onset of CIA in the optical is gradual, and its exact starting wavelength 
is difficult to quantify.

A cool white dwarf, as a companion to an M7 dwarf, might 
qualitatively explain the 
colors $J - K_{s}$, $J - W2$ of WISEA 0615$-$1247 (Figure~\ref{fig1}). CIA
models of cool white dwarfs 
have bumps and wiggles in the infrared, rather than a smooth flux
descent with increasing wavelength (D. Saumon and D. Koester, presented 
by Kilic et al.\ 2006; Abel et al.\ 2011). The $J - K_{s}$
color is similar to that in a Rayleigh-Jeans spectrum, while longer 
wavelength colors are redder. The $J - K_{s}$ color of WISEA 0615$-$1247
(Section~\ref{intro}) is $\sim$ 0.06 mag bluer than $J - K_{s} \sim$ 1.0 
for a typical M7 dwarf (Figure 14 of Kirkpatrick et al.\ 2008). The
$J - W2$ color of WISEA 0615$-$1247 (Section~\ref{intro}) is $\sim$ 0.18 mag
redder than $J - W2 \sim$ 1.5 for a typical M7 dwarf (Figure 7 of 
Kirkpatrick et al.\ 2011). We discuss these colors further at the
end of Section~\ref{white_dwarf_binaries}.

\subsubsection{Comparison Cool White Dwarfs} \label{cool_white_dwarfs}


We searched the literature for spectra of cool white dwarfs with 
which to fit the difference spectrum in Figure~\ref{fig7}. 
The cool white dwarfs we selected are 
SDSS J133001.13$+$643523.8, WD 2356$-$209, LHS 3250, 
WISEA 1702$+$7158B, 
and WD J0205$-$053. The first two show Na D absorption features,
while the last three do not, although they were considered as potential fits 
to the continuum of the difference spectrum in the blue optical.
Table 2 lists observational and physical characteristics of these five cool
white dwarfs, and we also describe their spectroscopic and other 
characteristics next.

\begin{itemize}

\item SDSS J133001.13$+$643523.8 (Harris et al.\ 2003)
is a He-atmosphere white dwarf whose optical 
spectrum exhibits metallic lines. Its Sloan Digital Sky Survey 
(SDSS; York et al.\ 2000) spectrum has prominent and very broad Na I D 
absorption near 5,892 \AA.
The SDSS spectrum, of very low signal-to-noise, tentatively exhibits 
Ca I absorption features at 4226, 4335, and 4455 \AA, 
and Ca II infrared triplet absorption features at 8498, 8542, and 8602 \AA\ 
(Harris et al.\ 2003). 
Unlike the
spectra of other DZ white dwarfs, that of SDSS J133001.13$+$643523.8 does 
not exhibit the 
Mg I triplet lines near 5,175 \AA, or the Ca II H \& K lines (3,968 and
3,934 \AA).

\item WD 2356$-$209 was discovered as a candidate
Galactic halo cool white dwarf by Oppenheimer et al.\ (2001).
It is a DZ-type white dwarf, and its optical spectrum shows a 
broad and prominent absorption feature from Na I D and possible 
absorption features from the Mg I triplet near 5,175 \AA\ 
(Oppenheimer et al.\ 2001). 
 
\item LHS 3250 (Luyten 1976) was discovered to be the first cool white 
dwarf with CIA by Harris et al.\ (1999). It is a DC-type white dwarf, 
meaning it has no discernible spectral features in the optical. 

\item LSPM J1702$+$7158N 
(L\'{e}pine \& Shara 2005)\footnote{The LSPM-North Catalog is accessible at \\ https://heasarc.gsfc.nasa.gov/W3Browse/all/lspmnorth.html.}  
is the secondary-component of a common-proper motion binary (10.8\arcsec\ 
separation); the primary 
component is LP 43-310 (Salim \& Gould 2003; L\'{e}pine \& Shara 2005). 
Kirkpatrick et al.\ (2016) rediscovered the common-proper motion system in 
an analysis of AllWISE high motion sources. Kirkpatrick et al.\ (2016) 
obtained a DBSP spectrum of LSPM J1702$+$7158N and they classified the 
object as a cool white dwarf.

\item WD J0205$-$053
was also discovered as a 
candidate Galactic halo cool white dwarf by Oppenheimer et al.\ (2001),
and was followed up through optical photometry and spectroscopy by 
Salim et al.\ (2004). No spectral lines are seen in the optical, hence its 
classification as DC-type. Salim et al.\ (2004) concluded it 
had a He atmosphere. However, the SED of a pure-He atmosphere white dwarf 
would be expected to be nearly identical to that of a blackbody 
(Kowalski \& Saumon 2006). Bergeron et al.\ (2005) obtained optical and
near-infrared ground-based photometry, and Kilic et al.\ (2009) obtained 
{\it Spitzer}
IRAC photometry of this and other similar objects. Kilic et al.\ (2009) 
fitted atmospheric models to these data, and WD J0205$-$053 itself was 
modeled with a pure-H atmosphere. Kilic et al.\ (2009) found that, in their
sample of 44 cool white dwarfs, only one had a pure-He atmosphere.

\end{itemize}

We also selected for comparison the white dwarf WD 2213$+$317, whose
coordinates in SIMBAD associate it with the AllWISE high motion object 
WISEA 221515.51$+$315731.9 (Kirkpatrick et al.\ 2016). 
Kirkpatrick et al.\ (2016) found that the colors $J - K_{s}$ = 0.889 $\pm$ 
0.025 mag and $J - W2$ = 1.081 $\pm$ 0.028 mag place the AllWISE object 
among normal M dwarfs in color space (Figure~\ref{fig1}). Kirkpatrick 
et al.\ (2016) pointed to other binary systems known from the literature, 
consisting of M and white dwarfs, that also fall at the location of normal 
M dwarfs, and suggested that WD 2213$+$317 was worthy of follow-up.
Unfortunately, the SIMBAD association of the white dwarf and the AllWISE
object is erroneous; the latter is an early-M type dwarf instead, as explained
in the Appendix.

\subsubsection{Binaries of Late-M and Cool White Dwarfs}\label{white_dwarf_binaries}

\begin{figure}
\includegraphics[angle=0, width=5.25in]{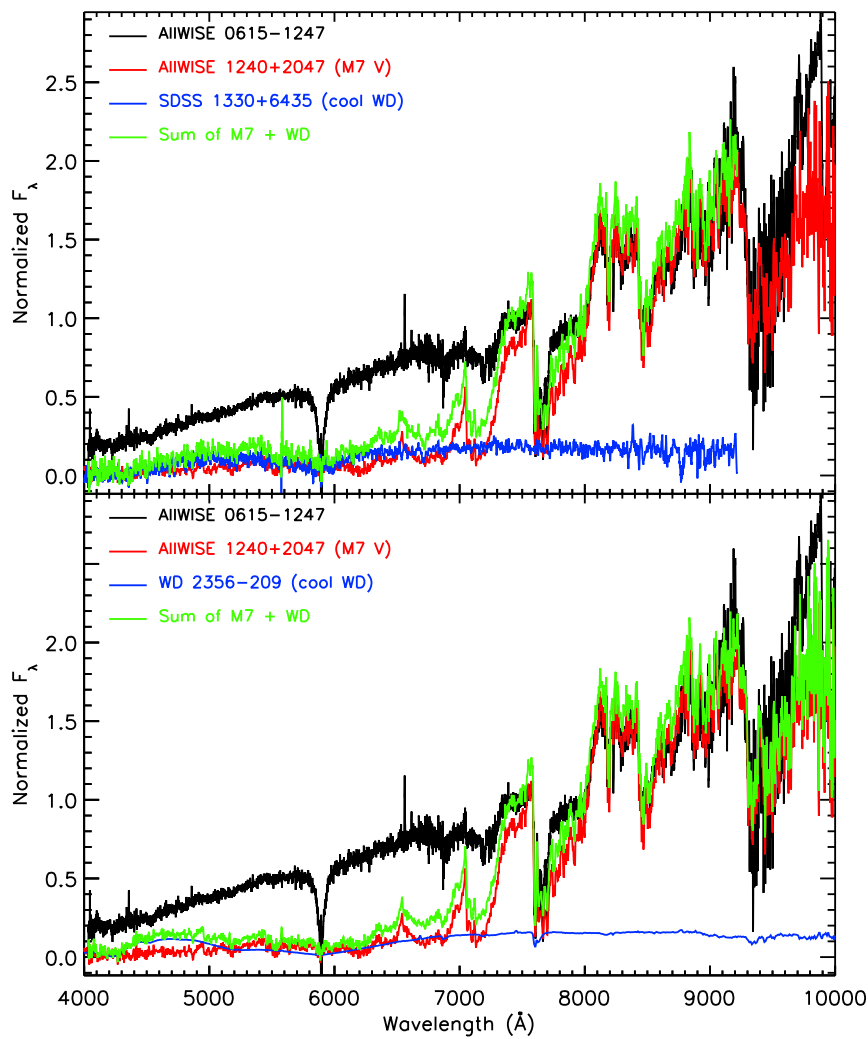}
\caption{\scriptsize Comparison of our optical (DBSP) spectrum of WISEA 0615$-$1247 from 
UT 2014 Oct 24 (shown in {\it black}), with the addition of an M7 dwarf
spectrum (shown in {\it red}, as in Figure~\ref{fig2}), and respectively
two cool white dwarf spectra (shown in {\it blue}). The 
{\it upper panel} includes the SDSS spectrum of SDSS J133001.13$+$643523.8 
(Harris et al.\ 2003), smoothed with a 5-pixel boxcar. The {\it lower panel} 
includes the LRIS spectrum (unpublished data courtesy S.\ Salim; see 
Salim et al.\ 2004 for details) of WD 2356$-$209 (Oppenheimer et al.\ 2001).
Both cool white dwarf spectra were separately normalized to 15\% of the 
observed 
7,400 \AA\ flux density of the WISEA 0615$-$1247 spectrum, and the M7 dwarf 
spectrum was normalized to the remainder (85\%) of this flux density. 
These empirical fractions were estimated as described in the text. The
addition of each cool white dwarf and M7 dwarf spectra, linearly 
interpolated to a common
wavelength grid, is shown in {\it green}. \label{fig8}}
\end{figure}

\begin{figure}
\includegraphics[angle=0, width=5.25in]{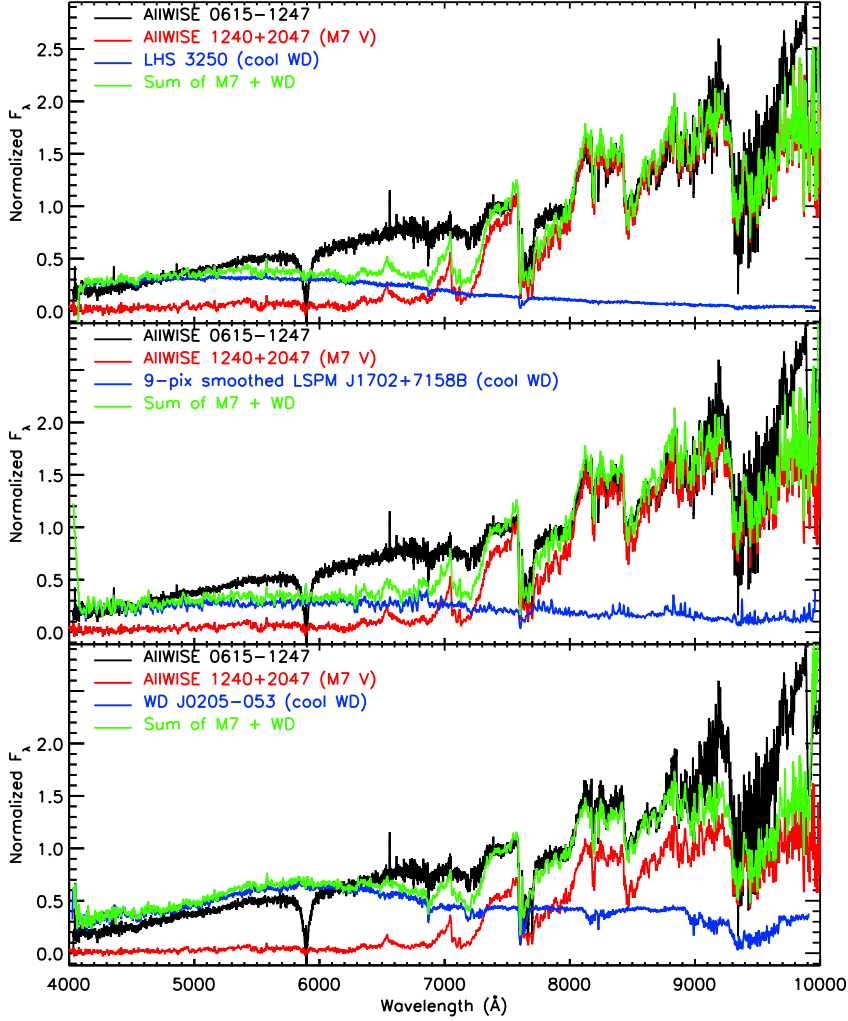}
\caption{\scriptsize Comparison of our optical (DBSP) spectrum of WISEA 0615$-$1247 from 
UT 2014 Oct 24 (shown in {\it black}), with the addition of an M7 dwarf
spectrum (shown in {\it red}, as in Figure~\ref{fig2}), and respectively
three cool white dwarf spectra (shown in {\it blue}). The 
{\it upper panel} includes our DBSP spectrum of LHS 3250 
(Harris et al.\ 2001). The {\it middle panel} includes the
Kirkpatrick et al.\ (2016) DBSP spectrum of
LSPM J1702$+$7158N (L\'{e}pine \& Shara 2005). The
{\it lower panel} includes our DBSP spectrum of WD J0205$-$053 
(Oppenheimer et al.\ 2001). The three cool white dwarf spectra were 
separately normalized to 15\%, 20\%, and 45\%, respectively, of the observed 
7,400 \AA\ flux density of the WISEA 0615$-$1247 spectrum, and the M7 dwarf 
spectrum was normalized to the remainder (85\%, 80\%, and 55\%, respectively) 
of this flux density. These empirical fractions were estimated as described
in the text. The addition of each cool white dwarf and M7 dwarf spectra, 
linearly interpolated to a common wavelength grid, is shown in 
{\it green}. \label{fig9}}
\end{figure}

Figures~\ref{fig8} and~\ref{fig9} show the comparison of our spectrum of 
WISEA 0615$-$1247 with spectral binary templates constructed from 
the above cool white dwarfs and an M7 dwarf. Spectra of the two cool 
white dwarfs with Na I D absorption listed in 
Section~\ref{cool_white_dwarfs}, SDSS J133001.13$+$643523.8 and 
WD 2356$-$209, are shown in 
Figure~\ref{fig8}, and spectra of the other three cool white dwarfs are 
shown in Figure~\ref{fig9}. The spectra of the binary 
components were each normalized at 7,400 $\mu$m. These fractions
were empirically determined by attempting to fit, after their addition, 
the blue optical spectral slope, the red optical detailed SED and spectral
features, and by avoiding flux densities in excess of the observed spectrum
of WISEA 0615$-$1247. After experimentation, we obtained fits using the
fractions listed in the captions of Figures~\ref{fig8} and~\ref{fig9}.

Figure~\ref{fig8} shows that the Na I D absorption in the two cool 
white dwarfs is 
substantially broader than in WISEA 0615$-$1247. It must be that 
pressure broadening is more pronounced in these cool white dwarfs, than
in our system. The spectral
energy distributions (SEDs) of these two cool white dwarfs are not
an adequate fit as binary companions in this case, because they produce
too much flux in the red optical, and not enough in the blue. Conversely,
the red spectrum is mostly fit by an M7 dwarf. CIA in these cool
white dwarfs appears to be less than in our system, possibly due to higher 
effective temperatures. We note that the temperature estimate of WD 2356$-$209
by Bergeron et al.\ (2005; see footnote (b) in Table 2) is relatively high.
However, the models by Bergeron et al.\ (2005) tend
to yield relatively high temperature values (H.\ Harris, priv.\ comm.).

In Figure~\ref{fig9} we show DBSP spectra for three cool white 
dwarfs; two
of them (LHS 3250 and WD J0205$-$053) were originally measured in the SDSS, 
but our new spectra have better signal-to-noise; the spectrum of the 
other 
source (LSPM J1702$+$7158N) was measured by Kirkpatrick et al.\ (2016).
Figure~\ref{fig9} shows that the spectra of the three cool white dwarfs are 
featureless
in the optical blue, thus not being adequate models for the Na I D 
absorption
in our system. The spectra of these three cool white dwarfs, however, 
have diminishing
flux with increasing wavelength in the optical red, likely from strong
CIA. These spectra contribute significantly to the hybrid spectrum in 
the optical blue. The above fits, however, are only 
approximately successful in the interval 
4,000--4,800 \AA\ for LHS 3250 (upper panel in Figure~\ref{fig9}), and 
4,000--4,600 \AA\ for LSPM J1702$+$7158N (middle panel in Figure~\ref{fig9}).
In particular, the slopes of the combined (binary) spectra (green lines in 
the upper and middle panels in Figure~\ref{fig9}) are slightly bluer than 
in the spectrum of our system (black lines in this Figure). The fit of 
the spectrum of WD J0205$-$053 (lower panel in Figure~\ref{fig9}), when
added to the same M7 dwarf, however, is quite promising.
This fit reproduces our spectral slope in the interval 
4,000--5,800 \AA, although with a small excess of flux.
Other limitations of this fit (green line in the bottom panel of 
Figure~\ref{fig9}) are a flux deficit at 6,200--7,300 \AA\ or, equivalently,
that the slope turnover due to CIA occurs at a redder wavelength in our
system, namely $\sim$ 6,800 \AA, than in WD J0205$-$053 ($\sim$ 6,000 \AA);
a deficit of flux in the optical red; and most notably lack of Na I D 
absorption. It is possible that the cool white dwarf in 
WISEA 0615$-$1247 has a higher abundance of Na than in WD J0205$-$053.

To check the consistency of our empirical relative flux normalizations of an 
M7 dwarf and a cool white dwarf, such as WD J0205$-$053, we first estimated 
synthetic broadband $V$ magnitudes from their spectra. We
then obtained absolute $V$ magnitudes $M_{V}$ of these objects from the 
literature, and 
compared the resulting relative fluxes. The spectra of the M7 dwarf and
WD J0205$-$053 in Figure~\ref{fig9}, when convolved with the Johnson-Morgan 
$V$ filter transmission function (Mermilliod, Mermilliod, \& Hauck 1997),\footnote{Johnson-Morgan-Cousins
$UVBRI$ filter transmission functions are in the \\ 
General Catalog of Photometric
Data, at http://obswww.unige.ch/GCPD/} indicate that the cool white dwarf is
a factor of $\sim$ 14 brighter than the M7 dwarf.\footnote{The synthetic $V$ magnitude we obtained from the spectrum of WISEA 0615$-$1247 is 20.4 mag, which differs by 0.8 mag from the estimate by L\'{e}pine (2008), listed in Table 1. The latter was derived from photographic magnitudes, using color-color models for field dwarfs, unreliable for an unusual binary system such as WISEA 0615$-$1247, and is subject to systematic errors and other effects (L\'{e}pine 2008).} 
On the other hand, an M7 dwarf has $M_{V} \sim$ 17.81 
(from an on-line tabular compilation\footnote{See http://www.pas.rochester.edu/$\sim$emamajek} referred to by Pecaut \& Mamajek 2013), 
while the cool white dwarf WD J0205$-$053 has 
$M_{V} \sim$ 16.59 (given its distance and apparent $V$ magnitude). 
Therefore, at a common distance from us, the cool white dwarf is predicted 
to be a factor of $\sim$ 3.1 brighter at $V$ band than an M7 dwarf. 
It is possible that 
the cool white dwarf in the WISEA 0615$-$1247 system is hotter than 
WD J0205$-$053, or has a larger radius, as discussed at the end of this
Section, for it to be so much brighter than the M7 dwarf.

An estimate of the photometric distance to WISEA 0615$-$1247 can 
be obtained in view that the cool white dwarf is comparatively much fainter 
than an M7 dwarf at $J$ band. An M7 dwarf 
has absolute magnitude $M_{J} \sim$ 10.8 (Hawley et al.\ 2002; 
on-line compilation$^{13}$ referred to by 
Pecaut \& Mamajek 2013), and WD J0205$-$053 
has $M_{J} \sim$ 14.4 (from its distance and 2MASS $J$), resulting
in a flux ratio of $\sim$ 27. Neglecting the contribution from the cool
white dwarf, the distance to our system is $\sim$ 58 pc, per the observed 
2MASS $J$ magnitude (Table 1).

The above distance to WISEA 0615$-$1247 and its proper motion estimated
from 2MASS and AllWISE (Section~\ref{obs_proper_motion}) imply a tangential
velocity of $\sim$ 165 km s$^{-1}$. We assessed the possibility that this
velocity suggested a subdwarf star in our system, even though our 
spectra are not consistent with a subdwarf (Section~\ref{single_object}). 
Traditionally, subdwarf
stars were considered to be in the galactic halo (Ryan \& Norris 1991; 
Gizis 1997; L\'{e}pine et al.\ 2003, 2007), and therefore dynamically 
``heated.'' The median tangential velocity in the galactic halo is 
$\sim$ 220 km s$^{-1}$ (Reid \& Hawley 2005), while the typical velocity in 
the galactic disk is $\sim$ 37 km s$^{-1}$ (Reid 1997). The escape velocity 
of the Galaxy near the Sun is $\sim$ 525 km s$^{-1}$ (Carney \& Latham 1987). 
In a recent study of the kinematics of 3,517 M-type objects of metallicity 
classes sdM, esdM, and usdM, Savcheva et al.\ (2014) 
found that the sdMs tend to be at the galactic thick or old disk, 
while the esdMs and usdMs are at the galactic halo. A histogram of the
tangential velocities of these objects showed a peak at $\sim$ 50 km s$^{-1}$,
and values almost reaching 600 km s$^{-1}$ (Savcheva et al.\ 2014). 
The tangential velocity of WISEA 0615$-$1247 is a factor of $\sim$ 4
times larger than in the galactic thin disk, and $\sim$ 3 times larger
than in the thick disk, although smaller than in the halo. The metallicity
of WISEA 0615$-$1247 is inconsistent with membership in the latter, but it 
is probable that it belongs to the thick disk. However, it should be
taken into account that there are high excursions in the velocities of 
individual objects around the above typical values.
Therefore, any definite conclusion on the galactic
population membership of WISEA 0615$-$1247 will require a detailed
kinematic study, including radial velocity and precise parallax
measurements.

We computed the $J - K_{s}$ and $J - W2$ colors of the binary 
system of
an M7 dwarf and the cool white dwarf WD J0205$-$053, to see if they would
reproduce the colors of WISEA 0615$-$1247 shown in Figure~\ref{fig1}. 
We used $J$ and $K$ photometry of WD J0205$-$053 by
Bergeron et al.\ (2005), and {\it Spitzer} IRAC channel 2 photometry by 
Kilic et al.\ (2009), which we regarded as equivalent to {\it WISE W2},
and computed absolute magnitudes. We used absolute magnitudes $M_{J}$ and
$M_{K}$ of an M7 dwarf from the on-line reference by 
Pecaut \& Mamajek (2013)$^{13}$. 
However, the fact that the
cool white dwarf is intrinsically much fainter than the M7 dwarf results
in minute changes of colors $J - K_{s}$ and $J - W2$ for this binary, 
relative to an M7 dwarf (Section~\ref{binary}), of $\sim$ 0.03 
and 0.05 mag, respectively.

A more luminous cool white dwarf would either have a higher effective
temperature, or a larger radius, than WD J0205$-$053. However, the effects
of these two possibilities on CIA have to be taken into account. 
Borysow et al. (1997) showed that CIA decreases with increasing effective 
temperature, and increases with surface gravity, and that CIA is 
pronounced for low metallicity (Z $<$ 0.1 Z$_{\odot}$) and effective 
temperature up to 4,000 K.
Bergeron et al.\ (2005) and Tremblay \& Bergeron (2008) computed pure-H
atmospheric models of cool white dwarfs. In these models, the color $V - H$
becomes redder by $\sim$ 0.10--0.17 mag when the effective temperature
increases from 4,200 K (as in models for WD J0205$-$053) to 4,500 K (the
temperature at which the above reddening effect, due to reduced CIA, is
reversed). In our attempts to fit the SED of WISEA 0615$-$1247 with binaries,
we assumed that most of the near-infrared flux is emitted by an M7 dwarf, 
hence constraining $V - H$ color variations to corresponding $V$ variations
in the cool white dwarf. It follows that an increase in effective temperature
has only a modest effect, and of opposite sense, in trying to solve the
discrepancy in the predicted and observed flux ratios of the M7 and the
cool white dwarf. In the models by Tremblay \& Bergeron (2008), the color
$V - J$ becomes redder by $\sim$ 0.05 mag when surface gravity log($g$)
decreases from 8.5 to 7.5, at an effective temperature of 4,500 K. The 
reddening effect is $\sim$ 0.1 mag at effective temperature of 4,200 K. It 
follows that an increase in radius of 
a cool white dwarf leads to only a modest reduction in CIA, if log($g$)
variations are within the above range. The
assumed surface gravity of cool white dwarf models of WD J0205$-$053 by 
Bergeron et al.\ (2005) and Kilic et al.\ (2009) is log($g$) = 8.0. Surface
gravity should remain close to this value, as a result of an increase in the
radius of a cool white dwarf, so that CIA is still significant.

A cool white dwarf of large radius but of surface gravity comparable
to that of WD J0205$-$053 may be hypothesized as a possibly improved fit,
as a binary component, to WISEA 0615$-$1247. 
Althaus et al.\ (2009)\footnote{See the on-line tabulation of models of white
dwarfs by Althaus et al.\ (2009) at: http://evolgroup.fcaglp.unlp.edu.ar/TRACKS/tracks\_heliumcore.html} modeled 
He-core, H-atmosphere white dwarfs, whose progenitors had supersolar 
metallicities. In these models, a 0.22 M$_{\odot}$ object with 
log(Z/Z$_{\odot}$) = 0.03, and effective temperature of 4,100--4,200 K, has 
a surface gravity of log($g$) = 7.12, $M_{V} \sim$ 15 mag, $B - V \sim$ 1.04 
mag, and $J - K \sim$ 0.2 mag. For comparison, the colors of WD J0205$-$053
from Bergeron et al.\ (2005) are $B - V$ = 1.32 mag and $J - K$ = 0.1 mag.,
not too diifferent from those of the above Althaus et al.\ (2009) model white
dwarf. A binary of this white dwarf and an M7 dwarf results in a shift
of $J - K$ towards the blue of $\sim$ 0.07 mag, relative to an M7 dwarf.
This shift is larger than for WD J0205$-$053, and comparable to that seen 
in WISEA 0615$-$1247 ($\sim$ 0.06 mag, Section~\ref{binary}).

\subsection{Other Possible Hypotheses}

Other scenarios we explored are:

\begin{itemize}

\item A triple system composed of two late-M dwarfs and a cool white dwarf. 
It is possible that a third component, in addition to the binary system we
discussed in Section~\ref{white_dwarf_binaries}, can improve the spectral fit
to our data, such as in covering the flux deficit we currently see in the 
interval 6,200-7,300 \AA. If this third component is an sdM, esdM, or
usdM, it would have to have been captured, because its low metallicity is 
disparate from the presence of TiO lines in our spectra. It is possible that
an sdM, esdM, or usdM can improve the spectral fit in the near-infrared.
Figure~\ref{fig5} shows that an M7 dwarf spectrum yields flux deficits at
9,700--11,000 \AA\ and 11,500-13,000 \AA, as well as a bluer slope
at 15,000--16,500 \AA, and/or a deficit at 16,800 \AA, relative to
WISEA 0615$-$1247. The relative flux contribution of an sdM, esdM, or usdM
in a multiple system, that would alleviate these shortcomings, is 
difficult to assess. However, as described at the end of 
Section~\ref{single_object}, an sdM, esdM, or usdM is not a good fit to the 
optical red spectrum of WISEA 0615$-$1247, and such remains the case
regardless of the flux ratios in a multiple system. The bulk of the 
flux contribution in the optical red would have to be from an M7 dwarf.

\item There could be unusual chromospheric activity in a single or multiple
stellar system. We would have expected emission
features, in contrast to the pronounced and broadened Na I D absorption
that we see. The UT 2014 Oct 24 spectrum exhibits modest H$_{\alpha}$ and
H$_{\beta}$ lines, that might indicate some level of this kind of activity.
However, white-light flare events in active M dwarfs, which occur due to
strong magnetic fields, produce much stronger H Balmer series lines than
seen in our object (Kowalski et al.\ 2010 and references therein), and there 
is no Na I D absorption.

\end{itemize}

\section{Conclusions}

We propose that WISEA 0615$-$1247 is an unresolved binary consisting of an
M7 dwarf and a cool white dwarf, the latter having an SED unlike any other 
cool white dwarf currently known. We speculate that the cool white dwarf 
produces the observed Na I D absorption feature, if Na abundance is 
high, and if pressure broadening is less than in other known cool white 
dwarfs. Additional studies of this fascinating 
system are warranted, including an accurate trigonometric parallax
measurement. 


\clearpage
\acknowledgments

We thank our referee, Derek Homeier, for suggestions and corrections 
that greatly improved the manuscript.
We also thank Hugh Harris and Samir Salim for very useful discussions that 
greatly benefited this paper, and Samir Salim for allowing us to 
use unpublished data.
This publication makes use of data products from {\it WISE}, which is a joint 
project of the University of California, Los Angeles, and the Jet Propulsion 
Laboratory (JPL)/California Institute of Technology (Caltech), funded by the 
National Aeronautics and Space Administration (NASA). This research has made 
use of the NASA/Infrared Processing and Analysis Center (IPAC) Infrared 
Science Archive, which is operated by JPL/Caltech, under contract with NASA. 
This publication also makes use of data products from the Two Micron All Sky 
Survey, which is a joint project of the University of Massachusetts and 
IPAC/Caltech, funded by NASA and the National Science Foundation (NSF).
The DENIS project has been partly funded by the SCIENCE and the HCM plans of
the European Commission under grants CT920791 and CT940627.
It is supported by INSU, MEN and CNRS in France, by the State of 
Baden-W\"urttemberg 
in Germany, by DGICYT in Spain, by CNR in Italy, by FFwFBWF in Austria, by 
FAPESP in Brazil, by OTKA grants F-4239 and F-013990 in Hungary, and by the 
ESO C\&EE grant A-04-046.
Funding for the SDSS and SDSS-II has been provided by the Alfred P. Sloan 
Foundation, the Participating Institutions, NSF, 
the U.S. Department of Energy, NASA, the Japanese Monbukagakusho, the 
Max Planck Society, and the Higher Education Funding Council for England. 
The SDSS Web Site is http://www.sdss.org/.
The SDSS is managed by the Astrophysical Research Consortium for the 
Participating Institutions. The Participating Institutions are the American 
Museum of Natural History, Astrophysical Institute Potsdam, University of 
Basel, University of Cambridge, Case Western Reserve University, 
University of Chicago, Drexel University, Fermilab, the Institute for 
Advanced Study, the Japan Participation Group, Johns Hopkins University, the 
Joint Institute for Nuclear Astrophysics, the Kavli Institute for Particle 
Astrophysics and Cosmology, the Korean Scientist Group, the Chinese 
Academy of Sciences (LAMOST), Los Alamos National Laboratory, the 
Max-Planck-Institute for Astronomy (MPIA), the Max-Planck-Institute for 
Astrophysics (MPA), New Mexico State University, Ohio State University, 
University of Pittsburgh, University of Portsmouth, Princeton University, 
the United States Naval Observatory, and the University of Washington.
The Digitized Sky Surveys were produced at the Space Telescope Science 
Institute under U.S. Government grant NAG W-2166. The images of these 
surveys are based on photographic data obtained using the Oschin Schmidt 
Telescope on Palomar Mountain and the UK Schmidt Telescope. The plates were 
processed into the present compressed digital form with the permission of 
these institutions.

\clearpage

\appendix

\section{Misidentification of the White Dwarf WD 2213$+$317 as \\ WISEA
221515.51$+$315731.9}\label{wd2213}

\begin{figure}
\includegraphics[angle=0, width=5.7in]{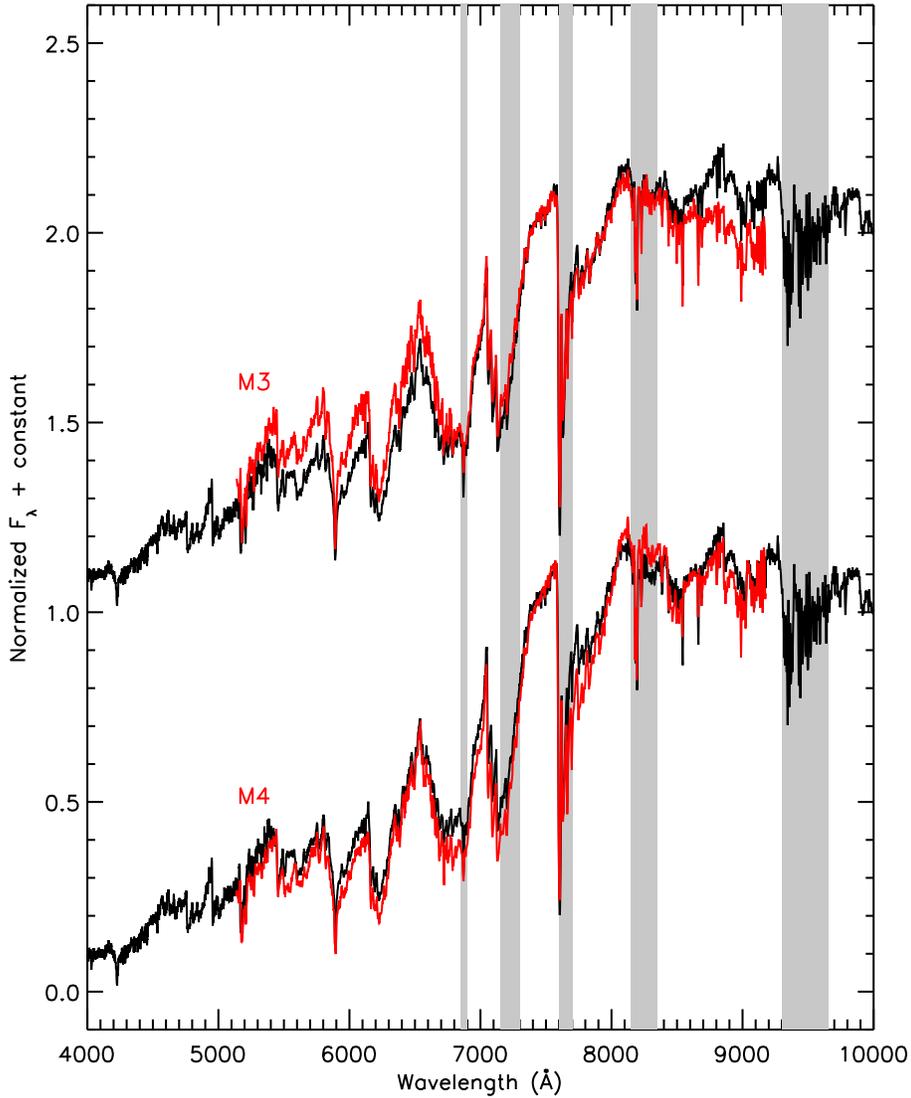}
\caption{\footnotesize The {\it black lines} are our DBSP spectrum of 
LP 343-35, rediscovered as the high motion source WISEA 221515.51$+$315731.9 
(Kirkpatrick et al.\ 2016). The spectrum is plotted 
in normalized flux density units, and was replicated vertically by adding 
offsets, for comparison with DBSP spectra of known M3 and M4 type 
dwarfs (shown as {\it red lines}). The object is of spectral type intermediate
between M3 and M4. Regions of low atmospheric transmission 
(Hamuy et al.\ 1994) are indicated by {\it gray bands}.\label{fig10}}
\end{figure}

The high motion object WISEA 221515.51$+$315731.9 is a rediscovery 
by Kirkpatrick et al. (2016) of LP 343-35, listed in Luyten's White Dwarf 
Catalogs (1970-1977)\footnote{Luyten's White Dwarf Catalogs (1970-1977) are 
accessible at \\ 
http://vizier.cfa.harvard.edu/viz-bin/VizieR-3?-source=III/70/catalog}
as a wide binary companion (more than 2 arcmin away) to LP 343-34. SIMBAD
erroneously associates the coordinates of the AllWISE object to LP 343-34,
which is the white dwarf WD 2213$+$317. Based on the SIMBAD association, Kirkpatrick
et al.\ (2016) suggested that the AllWISE object was an unresolved binary
of a white and an M dwarf, upon which we measured its spectrum with DBSP
(Figure~\ref{fig10}) to investigate its nature. Figure~\ref{fig10} also
shows comparison spectra of M3 and M4 dwarfs. This object is clearly an 
M3.5 dwarf.

We also investigated the putative binarity nature of LP 343-34 and 343-35.
The AllWISE Source Catalog lists the motion of LP 343-35 as RA motion = 
$-$54 $\pm$ 39 mas\, yr$^{-1}$ and Dec motion = $-$186 $\pm$ 35 
mas\, yr$^{-1}$. 
The displacement 
of the object between
its 2MASS and AllWISE positions yields $\mu_{\alpha}$ = 3.3 $\pm$ 6.4 
mas\, yr$^{-1}$
and $\mu_{\delta}$ = $-$111.8 $\pm$ 6.3 mas\, yr$^{-1}$ (Kirkpatrick et al.\ 2016), 
in rough agreement (within 3$\sigma$) with the AllWISE catalog motion. 
The object LP 343-34 = 2MASS J22150696$+$3158402 = LSPM 2215$+$3158
(L\'{e}pine \& Shara 2005) must be the white
dwarf WD 2213$+$317. Its listed proper motions in the LSPM-N Catalog are 
$\mu_{\alpha}$
= $-$47 mas\, yr$^{-1}$, and $\mu_{\delta}$ = $-$149 mas\, yr$^{-1}$ 
(L\'{e}pine \& Shara 2005). Luyten's
White Dwarf Catalogs (1970-1977) list a total motion of 128 
mas\, yr$^{-1}$ at P.A.
= 188$^{\circ}$, consistent with the above. Our measurement of 
proper motion from
the displacement between the 2MASS and AllWISE positions of LP 343-34
is $\mu_{\alpha}$ = $-$135.3 $\pm$ 29.8 mas\, yr$^{-1}$, and 
$\mu_{\delta}$ = $-$226.8 $\pm$ 31.8
mas\, yr$^{-1}$, in agreement, within 3$\sigma$, with the LSPM-N and Luyten's Catalogs.
The latter described LP 343-34 and 343-35 as a common proper motion binary.
However, the motions of the two objects are distinct, as can be seen by
comparing the 2MASS-to-AllWISE-derived values. Therefore, the white dwarf
WD 2213$+$317 and our discovered M3.5 dwarf are not physically associated.

\clearpage

\begin{deluxetable}{ccc}
\tabletypesize{\footnotesize}
\tablewidth{6.5in}
\tablecolumns{3}
\tablecaption{\sc General Observational Data of WISEA 0615$-$1247}
\startdata
\multicolumn{3}{l}{\sc Photometry} \\
\tableline
{\sc Bandpass} & {\sc Magnitude} & {\sc Error} \\
$V$ & 19.55\tablenotemark{a} & 0.50 \\ 
$J$ & 14.617\tablenotemark{b} &  0.031 \\
$H$ & 14.128\tablenotemark{b} &  0.034 \\
$K_{s}$ & 13.680\tablenotemark{b} & 0.049 \\
$W1$ & 13.294\tablenotemark{c} & 0.025 \\
$W2$ & 12.932\tablenotemark{c} & 0.026 \\
$W3$ & 12.462\tablenotemark{c} & 0.438 \\
$W4$ & 9.168\tablenotemark{c} (limit) & \nodata \\

\multicolumn{3}{l}{\sc Astrometry} \\
\tableline
{\sc Coordinates (J2000)} & {\sc Epoch} & {\sc Source} \\
06h 15m 43.5s\phantom{0}  $-$12d 47m 20s\phantom{.0}   & 1996.06 & DENIS \\
06h 15m 43.57s $-$12d 47m 22.1s & 1999.04 & 2MASS \\
06h 15m 43.5s\phantom{0}  $-$12d 47m 22s\phantom{.0}   & 1999.18 & DENIS \\
06h 15m 43.91s $-$12d 47m 26.8s & 2010.56 & AllWISE \\

\multicolumn{3}{l}{\sc Motions\tablenotemark{d}} \\
RA Motion (mas\, yr$^{-1}$) & Dec Motion (mas\, yr$^{-1}$) & {\sc Source} \\
593 $\pm$ 73 & $-$322 $\pm$ 78 & AllWISE \\
\multicolumn{3}{l}{\sc Proper Motion} \\
$\mu_{\alpha}$ (mas\, yr$^{-1}$) & $\mu_{\delta}$ (mas\, yr$^{-1}$) & {\sc Source} \\
452 $\pm$ 10 & $-$421 $\pm$ 10 & L\'{e}pine (2008) \\
436.9 $\pm$ 6.4 & $-$406.8 $\pm$ 6.5 & 2MASS-to-AllWISE \\
447 $\pm$ 5  & $-$414 $\pm$ 4  & DSS1 $R$, DSS2 $R$, 2MASS, \& {\it WISE} \\

\multicolumn{3}{l}{\sc Parallax} \\
$\pi$ (arcsec) & {\sc Error} (arcsec) & {\sc Source} \\
0.041 & 0.048 & DSS1 $R$, DSS2 $R$, 2MASS, \& {\it WISE} \\

\enddata
\tablenotetext{a}{From L\'{e}pine (2008), estimated from photographic magnitudes using the method of L\'{e}pine \& Shara (2005).}
\tablenotetext{b}{From 2MASS.}
\tablenotetext{c}{From AllWISE.}
\tablenotetext{d}{Motions from AllWISE are not strictly proper motions;
see the AllWISE Explanatory Supplement at 
http://wise2.ipac.caltech.edu/docs/release/allwise/expsup/.}

\end{deluxetable}

\clearpage

\begin{deluxetable}{lcccccc}
\tabletypesize{\footnotesize}
\tablewidth{6.5in}
\tablecolumns{7}
\tablecaption{\sc Characteristics of Comparison Cool White Dwarfs}
\tablehead{
\colhead{\sc Name} & \colhead{\sc Type} & \colhead{\sc Temperature} & \colhead{\sc Distance} & \colhead{\sc $\mu_{\alpha}$} & \colhead{\sc $\mu_{\delta}$} & \colhead{\sc Ref.\tablenotemark{a}} \\
\colhead{} & \colhead{} & \colhead{(K)} & \colhead{(pc)} & \colhead{(mas\, yr$^{-1}$)} & \colhead{(mas\, yr$^{-1}$)} & \colhead{} \\
\colhead{(1)} & \colhead{(2)} & \colhead{(3)} & \colhead{(4)} & \colhead{(5)} & \colhead{(6)} & \colhead{(7)} }
\startdata
SDSS J133001.13$+$643523.8 & DZ & \nodata & \nodata & $-$193 & 0 & 1 \\
WD 2356$-$209 & DZ & 3500--4500\tablenotemark{b} & 74 $\pm$ 34\tablenotemark{b} & $-$329 & $-$211 & 2 \\
LHS 3250 & DC & 3000--4000 & 30 & $-$548.1 $\pm$ 5.3 & 157.3 $\pm$ 5.3 & 3 \\
LSPM J1702$+$7158N & \nodata & \nodata & \nodata & $-$286.5 $\pm$ 13.1\tablenotemark{c} & 119.3 $\pm$ 7.3\tablenotemark{c} & 4 \\
WD J0205$-$053 & DC\tablenotemark{d} & 4200\tablenotemark{d} & 29 $\pm$ 8 pc & \multicolumn{2}{c}{1051 $\pm$ 25\tablenotemark{e}} & 2 \\
\enddata
\tablenotetext{a}{References to discoveries: (1) Harris et al.\ (2003). 
(2) Oppenheimer et al.\ (2001). (3) Harris et al.\ (1999); rediscovered by Harris
et al.\ (2001). (4) L\'{e}pine \& Shara (2005).}
\tablenotetext{b}{Distance and temperature estimates by Salim et al.\ (2004); 
Oppenheimer et al.\ (2001) estimate a distance of 85 pc; Bergeron et al.\ 
(2005) esimate a distance of 77 pc and a temperature of 4790 K.}
\tablenotetext{c}{2MASS-to-AllWISE proper motion estimates by 
Kirkpatrick et al.\ (2016). The motion from the AllWISE Catalog is 
RA motion = $-$303 $\pm$ 46 mas\, yr$^{-1}$ and Dec motion = 
7 $\pm$ 44 mas\, yr$^{-1}$, The proper
motion measured by L\'{e}pine \& Shara (2005) is $\mu_{\alpha} =$ $-$218 
mas\, yr$^{-1}$ and $\mu_{\delta} =$ 118 mas\, yr$^{-1}$).    }
\tablenotetext{d}{Spectral type and temperature estimates by 
Salim et al.\ (2004), very comparable to estimates by Bergeron et al.\ 
(2005) and Kilic et al.\ (2009).}
\tablenotetext{e}{The RA and Dec components of motion are not available
from Oppenheimer et al.\ (2001).}
\end{deluxetable}



\end{document}